\documentclass[showpacs,prl,10pt,a4paper,twocolumn]{revtex4}

\usepackage{graphicx}
\usepackage{times}
\usepackage{subfigure}
\usepackage{amssymb}
\usepackage{hyperref}

\begin{document}

\title{Generation of arbitrary Fock states via resonant interactions in
cavity QED}
\author{Gisana C. A. Bueno}
\affiliation{Instituto de F\'{\i}sica, Universidade Federal de Goi\'{a}s, 74.001-970, Goi%
\^{a}nia (GO), Brazil.}
\author{W. B. Cardoso}
\email[Electronic address.]{wesleybcardoso@gmail.com}
\affiliation{Instituto de F\'{\i}sica, Universidade Federal de Goi\'{a}s, 74.001-970, Goi%
\^{a}nia (GO), Brazil.}
\author{A. T. Avelar}
\affiliation{Instituto de F\'{\i}sica, Universidade Federal de Goi\'{a}s, 74.001-970, Goi%
\^{a}nia (GO), Brazil.}
\author{B. Baseia}
\affiliation{Instituto de F\'{\i}sica, Universidade Federal de Goi\'{a}s, 74.001-970, Goi%
\^{a}nia (GO), Brazil.}

\begin{abstract}
We propose a scheme to generate arbitrary Fock states $|N\rangle $ in a
cavity QED using $N$ resonant Rydberg atoms. The atom-field interaction
times are controlled via Stark-shifts adjusted in a way that each atom
transfers a photon to the cavity, turning atomic detections useless.
Fluctuations affecting the control of the atom-field interactions are also
considered.
\end{abstract}

\date{\today}
\maketitle

Fock states have various potential applications, as in secure quantum
communication \cite{Zbinden99,Gheri98,Enk97,Enk98}, quantum cryptography 
\cite{Jennewein00}, optimal capacity coding in quantum channels \cite%
{Caves94}, high-precision quantum interferometry \cite{Holland93}, etc.
However, their generation in laboratories is not a trivial task, mainly
concerning with highly excited fields. Recent experimental results for
one-photon \cite{Brattke01} and two-photon \cite{Bertet02} Fock states have
been obtained in a cavity QED taking advantage of the high level control of
the matter-field interaction \cite{Raimond01}. Proposals for the generation
of highly excited Fock states using a large number of atoms have been
presented \cite{Krause89,Brune90}. Ref. \cite{Krause89} employs a resonant
atom-field interaction and requires atomic detectors having high efficiency,
not available till now. The proposal in \cite{Brune90}, based on a
(dispersive) quantum nondemolition measurement of the photon number,
projects the cavity mode in an unpredictable Fock state. Alternative
approaches using superpositions of coherent \cite%
{Janszky94,Szabo96,Malbouisson99,Maia04,Aragao04,Avelar05a,Avelar05b} or
squeezed \cite{Monteiro05a,Monteiro05b} states distributed on a circle in
the phase space require a lesser number of atoms to generate Fock states ;
they also need highly efficient atomic detectors.

Pursuing the same goal, there are also proposals that employ a single atom
escaping the detection efficiency problem at price of complications in
atomic level schemes \cite{Law96,Santos01} or in successive atom-field
operations \cite{Domokos98}. In Ref.\cite{Law96} a three-level atom driven
by three classical fields via a two-channel Raman interaction transfers
photons to a cavity mode to prepare it in Fock states; in Ref.\cite%
{Domokos98} a simplified scheme prepares Fock states using a single
two-level atom which undergoes a controlled succession of interactions with
two modes of a cavity and transfer photons from one of them to the other.
The procedure in \cite{Domokos98} has a good accuracy and could achieve Fock
states $|N\rangle $ with $N\sim 5$.

In the present report, inspired by the work of Krause \textit{et al.} \cite%
{Krause89}, we present a scheme for generation of the arbitrary Fock states
in a cavity QED using resonant atom-field interactions. The underlying idea
is to send a set of Rydberg atoms with the same atomic velocity and
interaction times adjusted via Stark effect \cite{Domokos95} in such a way
that each atom transfers a photon to a cavity mode. In accord to the sudden
approximation \cite{Messiah58}, we will neglected the system evolution
between the active and frozen atom-field interactions. From the experimental
QED-cavity point of view this approximation is supported by the $1\mu s$
time switch spent by the atom between the electric fields 0.26 $V/cm$ and
1.1 $V/cm$ available in laboratories \cite{Rauschenbeutel00}. Hence, our
procedure differs from Ref.\cite{Krause89} which employs Rydberg \textit{ions%
} whose interaction times with the cavity field are determined by the
control of ionic velocities via an accelerating electric field. So, although
following the same fundamental idea, the present procedure differs from Ref.%
\cite{Krause89} in the control of the atom-field interaction times.

The simplicity of our scheme makes it attractive experimentally, being
feasible with the present status of QED-cavity technology \cite%
{Gleyzes_NAT07}. The proposal requires the experimental setup shown in Fig.
1: the source $S$ ejects rubidium atoms, which are velocity selected and
prepared in circular Rydberg state, one at a time, by appropriated laser
beams. The relevant atomic levels $\left\vert g\right\rangle $ and $%
|e\rangle $, with the principal quantum numbers 50 and 51, have the atomic
transition of 51.1 GHz. After \textit{S} one obtains a known atomic position 
$r\left( t\right) $ at any time during the experiment. The high-Q
superconducting cavity $C$ is a Fabry-Perot resonator made of two spherical
niobium mirrors with a Gaussian geometry (waist $w=6~mm$) and photon damping
times of $130~ms$ \cite{Gleyzes_NAT07}. The cavity is prepared at a low
temperature ($T\simeq 0.6$ K) to reduce the average number of thermal
photons; before the beginning of the experiment the thermal field is erased; 
$D_{e}$ $(D_{g})$ represents the atomic ionization detector for the state $%
|e\rangle $ $($ $|g\rangle $ $)$.

To describe atom-field interaction in the cavity we employ the
Jaynes-Cummings model \cite{Jaynes63}, also including variation of the
strength of the local atom-field coupling. In the
rotating-wave-approximation this model is represented by the Hamiltonian, 
\begin{equation}
\hat{H}=\frac{\hbar \omega _{0}}{2}\hat{\sigma}_{z}+\hbar \omega \hat{a}%
^{\dagger }\hat{a}+\hbar \Omega (t)(\hat{a}^{\dagger }\hat{\sigma}_{-}+\hat{a%
}\hat{\sigma}_{+}),  \label{HT}
\end{equation}%
where $\hat{\sigma}_{z}=|e\rangle \langle e|-|g\rangle \langle g|$, $\hat{%
\sigma}_{+}=|e\rangle \langle g|$, and $\hat{\sigma}_{-}=|g\rangle \langle
e| $ are the atomic operators of the two-level atom with the atomic
transition frequency $\omega _{0}$; $\hat{a}$ ($\hat{a}^{\dagger }$) is the
annihilation (creation) operator of the single-mode field of frequency $%
\omega $; $\Omega (t)$ is the atom-cavity interaction strength with a
Gaussian mode profile \cite{Raimond01} 
\begin{equation}
\Omega (t)=\Omega _{0}\exp \left[ -\frac{r^{2}(t)}{w^{2}}\right] ,
\label{Gaussian}
\end{equation}%
where $\Omega _{0}$ stands for the vacuum Rabi oscillation at the center of
the cavity and the atomic position is described classically, $r\left(
t\right) =r_{0}+vt,$ since the kinetic energy of an atom is much larger than
the height and the depth of the optical potential \cite{Breuer01}. The
effective resonant Hamiltonian for the atom-field system in the interaction
picture is 
\begin{eqnarray}
\hat{V}_{JC}(t)&=&\hbar \Omega (t)(\hat{a}^{\dagger }\hat{\sigma}_{-} + \hat{a} \hat{\sigma}_{+}). 
\end{eqnarray}
Since the time dependence of this Hamiltonian comes from a parameter, $%
\Omega (t)$, then $[\hat{V}_{JC}(t),\hat{V}_{JC}(t^{\prime })]=0$ and \ \
the time evolution operator has the form $\hat{U}_{JC}(t)=exp[(-i/\hbar
)\int_{0}^{t}\hat{V}_{JC}(t^{\prime })dt^{\prime }].$ Thus, considering the
cavity mode in the Fock state $|n-1\rangle $ and the atom in the excited
state $|e\rangle $, this evolution operator $\hat{U}_{JC}$ produces 
\begin{eqnarray}
\hat{U}_{JC}(\tau _{n})|n-1\rangle |e\rangle &=&\cos (\sqrt{n}\theta (\tau
_{n}))|n-1\rangle |e\rangle \nonumber \\ &-& i\sin (\sqrt{n}\theta (\tau _{n}))|n\rangle
|g\rangle ,  \label{evol}
\end{eqnarray}
with 
\begin{equation}
\theta (\tau _{n})=\int_{0}^{\tau _{n}}\Omega (t)dt,  \label{time}
\end{equation}%
where $\tau _{n}$ stands for the atom-field interaction time concenrnig with
the $n$-$th$ atom.

\begin{figure}[t]
\centering
\includegraphics[{width=7cm}]{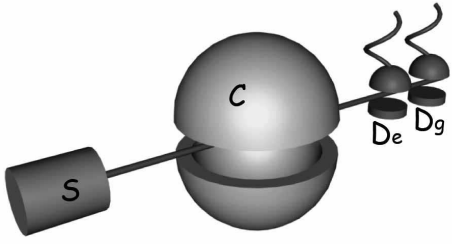}
\caption{Scheme of the setup creating $|N\rangle$.}
\end{figure}

Now, to describe our procedure to generate arbitrary Fock states consider a
first atom prepared in the excited state $|e\rangle_{1}$, which enters the
cavity and interacts resonantly with a field mode in a vacuum state $%
|0\rangle$. According to the Eq. (\ref{evol}) the initial state $%
|0\rangle|e\rangle_{1}$ evolves to the state $|1\rangle|g\rangle_{1}$ after
an interaction time $\tau_{1}$ obtained from the Eq. (\ref{time}) plus the
condition $\theta (\tau_{1})=\pi /2.$

Next, we send a second atom prepared in state $|e\rangle _{2}$ which
interacts with the field in the state $|1\rangle$ obtained in the previous
step. During the interaction time $\tau _{2}$ obtained from the condition $%
\theta (\tau _{2})=\pi /(2\sqrt{2})$ the atom-field system evolves to the
state $|2\rangle|g\rangle_{2}$. Proceeding further in this way, after the
passage of the $N$-$th$ atom one obtains the desired Fock state inside the
cavity 
\begin{equation}
|\psi (\tau _{N})\rangle _{AF}=|N\rangle |g\rangle_{N} ,
\end{equation}%
where $\tau_{N}$ stands for the interaction time in last step.

To calculate the interaction time for the creation of the Fock state $%
|N\rangle $ we take typical values \cite{Rauschenbeutel00,Raimond01} for the
coupling constant $\Omega _{0}\simeq 2\pi \times 47~kHz$ and velocity $%
\upsilon =500~m/s$ for all atoms. Using the Stark effect one can ajust all
interaction times in such a way that each atom has $100\%$ probability for
emitting a photon inside the cavity. For example, when the first atom enters
the cavity the atom-field interaction is frozen by a $1.1~V/cm$ field. When
the atom is $1.4~mm$ before the cavity axis, it is rapidly tuned in
resonance with cavity mode by a $0.26~V/cm$ field. During the next $5.4~\mu
s $ ($2.8~mm$ path) the atom emits a photon to the cavity. After that the
atom-field interaction is canceled out again. The same procedure is repeated
for all atoms. The instant of an atom entering the cavity coincides with
that of the previous atom exiting the cavity, so the total interaction time
is $N\overline{\tau }=N(l/\upsilon )$, $l$ being the length of the cavity.
For example, assuming $\overline{\tau }=10$ $\mu s$ the creation of the
number state $|6\rangle $ requires the total interaction time $60~\mu s$,
much lesser than the decoherence time $t_{d}=t_{cav}/N\simeq 12.3~ms$, with
damping time $t_{cav}\simeq 123$ $ms$ \cite{Gleyzes_NAT07}. So the scheme is
experimentally feasible within the realm of microwave.

Since the present scheme involves no atomic detection, in the ideal case the
desired Fock state is obtained with $100\%$ succes rate and fidelity.
However, to be more realistic we have taken into account variations of the
atomic velocity coming from the size of the excitation laser beams and the
residual velocity dispersion \cite{Raimond01}. Nowadays, the best accuracy
in the variance of the velocity is $\Delta v=\pm $ $2m/s$ which yields
atomic position with $\pm $ $1$ $mm$ accuracy. Also, there is no fundamental
problem to get a more accurate velocity and sufficiently well known atomic
position, since even improving the accuracy in the velocity for $\Delta v=$\ 
$\pm $ $2\times 10^{-3}m/s$ the Heisenberg uncertainty would furnish $\Delta
x\simeq \hslash /m\Delta v\simeq 0.35\mu m$, $m$ standing for rubidium mass.
Accordingly, the uncertainty in the atomic velocity leads to the
impossibility of sharply fixing the atom-field interaction times. Following
the Ref. \cite{Messina02} we introduce the probability density $f_{i}(t_{i},%
\tilde{t}_{i})$ where $t_{i}$ is the true atom-field interaction time. For a
Gaussian distribution centered around the average interaction time $\tilde{t}%
_{i}$, the probability density reads 
\begin{equation}
f_{i}(t_{i},\tilde{t}_{i})=\frac{1}{\Delta _{i}\sqrt{2\pi }}\exp \left\{ -%
\frac{(t_{i}-\tilde{t}_{i})^{2}}{2\Delta _{i}^{2}}\right\} ,  \label{p1}
\end{equation}%
where $\Delta _{i}=\gamma \tilde{t}_{i}$. The spread parameter $\gamma $
characterizes the control of the atom-field interaction time and it usually
fluctuates from $0$ to $0.1$ \cite{Messina02}. When $\Delta _{i}\rightarrow
0 $ the function $f_{i}(t_{i},\tilde{t}_{i})$ becomes a Dirac distribution $%
\delta (t_{i}-\tilde{t}_{i})$ corresponding to the ideal control of the
atom-field interaction time. The effective density operator $\tilde{\rho}%
_{AF}^{N},$ which describes the whole atom-field state during the generation
of the Fock state $|N\rangle $, including unavoidable influences of
fluctuations upon the interaction time, may be represented as%
\begin{equation}
\widetilde{\rho }_{AF}^{N}=\prod_{i=1}^{N}\int_{-\infty }^{+\infty
}dt_{i}f_{i}(t_{i},\tilde{t_{i}})\rho _{AF}^{N}(t_{1},\ldots ,t_{N}),
\label{p2}
\end{equation}%
where $\widetilde{t_{i}}=\pi /(2\sqrt{i}\lambda )$, $i=1,2,\ldots N$ and 
\begin{eqnarray}
\rho _{AF}^{N} (t_{1},\ldots ,t_{N})&=&\hat{U}_{JC}(t_{N})\hat{U}%
_{JC}(t_{N-1})\ldots \nonumber \\ &\times& \hat{U}_{JC}(t_{1})\hat{\rho}_{AF}(0)\hat{U}%
_{JC}^{\dagger }(t_{1})\ldots \nonumber \\ &\times& \hat{U}_{JC}^{\dagger }(t_{N-1})\hat{U}%
_{JC}^{\dagger }(t_{N}),
\end{eqnarray}%
with 
\begin{equation}
\hat{\rho}_{AF}(0)=|e\rangle _{1}|e\rangle _{2}\ldots |e\rangle
_{N}|0\rangle _{F}{}_{F}\langle 0|{}_{N}\langle e|\ldots {}_{2}\langle
e|{}_{1}\langle e|.
\end{equation}

Now, to obtain the success rate $P$ and the fidelity $F$, considering
fluctuations affecting the atom-field interaction time, we use the
definitions 
\begin{equation}
P=\left\{ \prod\limits_{i=1}^{N}\left[ _{i}\left\langle g\right\vert \right]
_{c}\left\langle N\right\vert \right\} Tr_{F}[\widetilde{\rho }%
_{AF}^{N}]\left\{ \prod\limits_{j=1}^{N}\left[ \left\vert g\right\rangle _{j}%
\right] \left\vert N\right\rangle _{c}\right\}
\end{equation}%
and 
\begin{equation}
F=\langle N|Tr_{A}[\widetilde{\rho }_{AF}^{N}]|N\rangle ,
\end{equation}%
where $Tr_{F}$ and $Tr_{A}$ stand for the trace on the field and atomic
states, respectively. Here the success rate and fidelity coincide and they
are given by 
\[
F(\gamma ,N)=\left[ \left( 1+e^{-\frac{\gamma ^{2}\pi ^{2}}{2}}\right) /2%
\right] ^{N}. 
\]%
Note that, as expected, for $\gamma =0$ corresponding to the ideal case, the
fidelity is $100\%$. Fig. 2 shows the fidelity of the Fock state $|N\rangle $
obtained in the present scheme, for the interval $N\in \lbrack 1,10]$.
Accordingly the minimum value of fidelity is $79\%$, for $\gamma =0.1 $, $%
N=10$.

\begin{figure}[t]
\centering
\includegraphics[width=7cm]{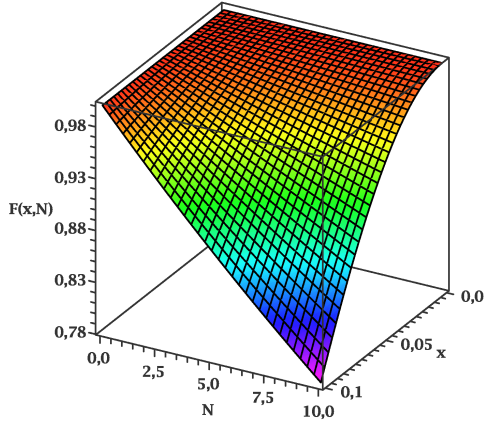}
\caption{Fidelity of Fock state $|N\rangle$ considering the fluctuations
effects in the atom-field interaction time; $x=\gamma$.}
\end{figure}

In summary, we have employed a set of $N$ Rydberg atoms to create an
arbitrary Fock state $|N\rangle $ inside a microwave cavity via resonant
atom-field interactions. The variation of the Rabi frequency due to the
atomic motion across the Gaussian cavity mode was taken into account to
calculate the generation time; for example, it takes $0.2$ $ms$ to create
the number state $|6\rangle $ with the success probability of $100\%$ in the
ideal case. We have also investigated the loss of fidelity (cf. Fig. 2) due
to the variation of the atomic velocity caused by the size of the excitation
laser beams and the residual velocity dispersion \cite{Raimond01}. Finally,
a recent result by the Haroche's group \cite{Gleyzes_NAT07} allows us to
neglect the decoherence of the state during the generation process.

\section*{Acknowledgments}

The partial support from the CNPq, Brazilian agency, is gratefully
acknowledged.


\end{document}